\documentclass[aps,prd,twocolumn,superscriptaddress,floatfix,showpacs]{revtex4-1}
\usepackage{amsmath, amsfonts}
\usepackage{graphicx}

\newcommand{\mnras}{Mon.\ Not.\ R.\ Astro.\ Soc.}
\newcommand{\aj}{Astron. J.}
\newcommand \nubar {\bar{\nu}}
\newcommand \ybar {\bar{y}}

\newcommand \beq {\begin{equation}}
\newcommand \eeq {\end{equation}}
\newcommand \beqn {\begin{eqnarray}}
\newcommand \eeqn {\end{eqnarray}}
\newcommand \apjl {ApJL}

\begin{document}
\title{Accretion Disks Around Binary Black Holes: A Quasistationary Model}
\author{Yuk Tung Liu} 
\author{Stuart L. Shapiro}
\altaffiliation{Also Department of Astronomy and NCSA, University of
  Illinois at Urbana-Champaign, Urbana, IL 61801}
\affiliation{Department of Physics, University of Illinois at
  Urbana-Champaign, Urbana, IL 61801}

\begin{abstract}
Tidal torques acting on a gaseous accretion disk around a 
binary black hole can create a gap in the disk near the 
orbital radius.
At late times, when the binary inspiral timescale due to 
gravitational wave emission becomes shorter than
the viscous timescale in the disk, the binary 
decouples from the disk and eventually merges. 
Prior to decoupling the balance between tidal and 
viscous torques drives the disk to a quasistationary equilibrium state, 
perturbed slightly by small amplitude, spiral density waves emanating 
from the edges of the gap.  We consider a black hole binary with a 
companion of smaller mass and construct a simple Newtonian
model for a geometrically thin, Keplerian disk in 
the orbital plane of the binary. We solve the disk evolution equations 
in steady state to determine the quasistationary, (orbit-averaged)  surface 
density profile prior to decoupling. 
We use our solution, which is analytic up to simple quadratures,
to compute the electromagnetic flux and approximate radiation spectrum
during this epoch. A single nondimensional parameter $\tilde{g}$, equal to 
the ratio of the tidal to viscous torque at the orbital radius, determines 
the disk structure, including the surface density profile, the extent 
of the gap, the existence of an inner disk, and the accretion rate. 
The solution reduces to the Shakura-Sunyaev
profile for a stationary accretion disk around a single black hole in the
limit of small $\tilde{g}$.
Our solution may be useful for choosing physical parameters 
and setting up quasistationary disk initial data for detailed numerical simulations that
begin prior to decoupling and track the subsequent evolution of 
a black hole binary-disk system.

\end{abstract}
\maketitle

\section{Introduction}

Binary black hole (BHBH) mergers typically occur in regions immersed
in gas, and the capture and accretion of the gas by the binary may result in 
appreciable electromagnetic radiation. Following the detection of gravitational
waves from a BHBH merger, electromagnetic ``afterglow'' 
radiation could provide confirmation of the 
coalescence~\cite{MilP05, RosLAPK09, SchK08, CorHM09, OneMBRS09, Sha10, 
TanM10}. Such electromagnetic radiation can also serve as a useful probe
of the gas in galaxy cores or in other regions where mergers take place,
as well as a diagnostic of the physics of black hole accretion. The
timescale during which detectable ``afterglow'' radiation achieves its maximum
value occurs when the gas is driven close to the remnant and 
ranges from several years to tens of decades in the case of supermassive 
BHBH systems with total masses of $10^5 - 10^8 M_{\odot}$. 
Together with detecting the gravitational waves, 
observing this electromagnetic radiation may even provide a means of 
witnessing the birth of a quasar~\cite{TanHM10}.

There is also the possibility of detecting electromagnetic ``precursor''
radiation prior to the merger and before the maximum gravitational wave 
emission~\cite{ArmN02,ChaSMQ10}.  If the distant gas is nearly 
homogeneous and either at rest with respect to the binary
(``binary Bondi'' accretion) or moving 
(``binary Bondi-Hoyle-Lyttleton'' accretion)
and optically thin, the luminosity will peak at the end of the 
binary inspiral phase immediately prior to the final 
plunge~\cite{FarLS10}. At this stage  shock heating of 
the gas and turbulent magnetic field amplification is strongest. 
The peak luminosity lasts for 
$\delta t \sim M_6$ hours prior to merger and then
plummets sharply following the coalescence.  Here $M_6$ is the binary mass 
in units of $10^6 M_{\odot}$. If, instead, the accretion takes place via
a geometrically-thin, optically-thick Keplerian disk around the binary
(``binary Shakura-Sunyaev'' accretion), there may be a late-time 
precursor brightening from tidal and viscous (or turbulent magnetic) 
dissipation in the inner disk. This radiation peaks on a 
timescale $\delta t \sim 0.1 M_6$ days
prior to merger and it remains high afterwards~\cite{ChaSMQ10}.

In this paper we focus on geometrically thin disks prior to 
disk-binary decoupling and well before any late-time
brightening of the ``precursor'' electromagnetic radiation. 
Our calculations are based on a simplified, Newtonian prescription 
for a Keplerian disk in the orbital plane of a binary BHBH system with a low mass ratio. 
Similar disk equations have been integrated previously in time 
to follow the (secular) evolution of such a BHBH-disk system 
for selected cases (see, e.g.,~\cite{ArmN02, ChaSMQ10}). 
Here we adopt the equations 
to solve for steady state. We then apply our solution 
to determine the orbit-averaged disk structure and electromagnetic radiation 
during the inspiral epoch prior to decoupling, when our quasistationary 
approximation is appropriate.

Our simple treatment determines the quasistationary, 
orbit-averaged, surface density profile $\Sigma(r)$ of the 
circumbinary disk prior to 
decoupling, as well as the accretion rate, 
luminosity and approximate spectrum of the electromagnetic radiation. 
Our semi-analytic analysis, (i.e.\ analytic up to simple quadratures), 
serves to identify some of the key physical 
parameters that determine these quantities. Several nondimensional 
parameters fix the overall shape of the density profile.
Among these parameters are the binary mass ratio $q$, assumed small, the 
ratio of the disk scale height to radius, $h/r$, also small,
in the ring-like gap in the disk at the orbital radius of the 
secondary, and the power-law $n$ defining the 
variation of the disk viscosity with radius, $\nu \propto r^n$. Other 
parameters, such as the total binary mass $M$,  determine the
physical scale of the disk and, together with the accretion rate
$\dot M$, the characteristic luminosity and frequency of the emitted
radiation.

We find that many of the properties of the disk, such as the surface density 
profile,  
the extent of the gap at the orbit of the secondary, the existence of 
an inner disk and the accretion rate, are determined by a single nondimensional parameter $\tilde{g}$. 
This parameter is essentially the ratio of the tidal to viscous torque 
at the orbital radius.

A reliable description of the accretion flow and associated
radiation from a merging BHBH binary really requires a  
radiation magnetohydrodynamics simulation in full general 
relativity in a  $3+1$-dimensional, dynamical spacetime. Such 
simulations have yet to be performed in any detail. However, 
Newtonian hydrodynamic 
simulations incorporating some of the relevant physics have been performed at 
various levels of approximation 
(see, e.g.,~\cite{MacM08, CorHM09, RosLAPK09, OneMBRS09})
and general relativistic simulations are underway 
(e.g.,~\cite{MegAFHLLMN09,AndLMN10, FarLS10, BodHBLS10,BogBHLS10,FarLS11,MostraPRLYP10,ZanottiRDP10,PalenzuelaGLL10}).
The model discussed here, although based on a simplified description,
can help select input parameters and identify scaling behavior 
for such simulations. In addition, the resulting quasistationary 
description of the disk provides approximate initial data for 
numerical simulations that begin prior to binary-disk decoupling.

We adopt geometrized units and set $G=1=c$ below.

\section{Basic Model}
\subsection{Timescales and Overview}

There are several characteristic  timescales that determine the structure
of a gaseous disk in a BHBH system. The orbital timescale in a Keplerian 
disk with orbital angular frequency $\Omega_{K}(r) = (M/r^3)^{1/2}$ is given by
\begin{equation}
\label{eq:torb}
 t_{\rm orb}(r) = 2 \pi \left( \frac{r^3}{M} \right)^{1/2}\ ,
\end{equation}
where $r$ is local disk radius. The viscous timescale is given by 
\beq
\label{eq:tvis}
t_{\rm vis}(r) = \frac{\Sigma}{(d\Sigma/dt)_{\rm vis}} 
   \approx  \frac{M_{\rm disk} r^2 \Omega_K(r)}{T_{\rm vis}} \approx
\frac{2}{3}\frac{r^2}{\nu}\ , 
\eeq
where $\nu$ is the shear viscosity, $T_{\rm vis}$ is the viscous 
torque and $M_{\rm disk} \sim r^2 \Sigma(r)$ is the disk mass. 
The timescale associated with the 
gravitational tidal torque on the disk due to the companion is
\beq
  t_{\rm tid}^{\rm disk}(r) = \frac{\Sigma}{(d\Sigma/dt)_{\rm tid}}
\approx \frac{M_{\rm disk} r^2 \Omega_K(r)}{T_{\rm tid}} ,
\eeq
where $T_{\rm tid}$ is the gravitational tidal torque. The inspiral 
timescale of the binary due to the emission of gravitational waves is
\begin{equation}
\label{eq:GW}
t_{\rm GW}(a) = \frac{a}{(da/dt)_{\rm GW}} = \frac{5}{16} \frac{a^4}{M^3 \zeta}\ ,
\end{equation}
where $a$ is the binary separation and $\zeta \equiv 4q/(1+q)^2$. Finally, 
the binary orbit is also changed by the torque exerted by the disk on the 
secondary of mass $m=qM$. The timescale is 
\beq
  t_{\rm tid}^m = \frac{a}{(da/dt)_{\rm tid}} \approx
\frac{m a^2\Omega_K(a)}{T_{\rm tid}} 
\approx \frac{m}{M_{\rm disk}} t_{\rm tid}^{\rm disk}(a) .
\label{eq:ttidm}
\eeq

The inequality
$t_{\rm orb}(r) \ll t_{\rm vis}(r)$ is satisfied throughout the disk at all
times.  For most of its history, the disk evolves on a slow, secular,  
viscous timescale and not on a rapid dynamical timescale~\cite{DYN}; 
viscosity maintains the gas in nearly Keplerian circular orbits. 
In a quasistationary state, where the effects of viscosity and tidal 
torques balance each other, we have 
$t_{\rm tid}^{\rm disk} \sim t_{\rm vis}$. Consider the typical case 
where the disk mass satisfies $M_{\rm disk} \ll m$. It follows from Eq.~(\ref{eq:ttidm}) 
that $t_{\rm tid}^{\rm disk}(a) \ll t_{\rm tid}^m$. Hence, as long as the 
disk is quasistationary, we have $t_{\rm vis}(a) \ll t_{\rm tid}^m$.  
However, $t_{\rm tid}^m$ can be longer or shorter 
than $t_{\rm GW}$ during the early inspiral history. Prior to decoupling, 
even as the binary separation shrinks due to the combined effects of gravitational 
radiation and tidal torque back-reaction, the orbital 
separation always remains momentarily 
``frozen'' while the disk adjusts to tidal-viscous torque 
balance to maintain quasiequilibrium. Eventually, as the binary inspirals further, 
the timescale due to gravitational wave emission becomes shorter than the 
viscous timescale and the binary decouples from the 
disk~\cite{ArmN02,LiuWC03,MilP05}. We note that as long as 
$M_{\rm disk}/m \ll 1$, $t_{\rm vis} \ll t_{\rm tid}^m$ holds. 
Hence at decoupling and thereafter, the orbital decay is driven 
by gravitational radiation, not the torque of the disk. Quasiequilibrium 
no longer applies after decoupling.

The orbital 
radius at which decoupling begins can be estimated from the relation
$t_{\rm GW}(a) \sim \beta t_{\rm vis}(2 \lambda a)$. Here
$\lambda \equiv r/2a \sim 1$ is the nearly constant ratio between the 
disk edge and the orbital separation prior to decoupling and is 
determined by a balance between viscous stresses in the disk and
gravitational tidal torques from the binary ~\cite{ArtL94, MacM08}. 
The parameter $\beta \sim 0.1$ roughly accounts for the shortening 
of the viscous timescale 
at the edge where the  surface density $\Sigma$ is very steep ~\cite{LynP74}. 
An $\alpha$-disk with a viscosity law 
$\nu(r) = (2/3) \alpha P_{\rm gas}/(\rho \Omega_{\rm K})$, where 
$\rho$ is the gas density and $P_{\rm gas}$ is the gas pressure, 
yields a decoupling radius $a_d$ given by ~\cite{MilP05,TanM10}
\begin{equation}
\label{eq:ad}
\frac{a_{\rm d}}{M}  \approx 126  \alpha_{-1}^{-17/50} S^{-49/200} 
\lambda ^{7/10} M_6^{2/25}(\beta_{-1} \zeta)^{17/40} \theta_{0.2}^{-17/200}\ ,
\end{equation}
where $\alpha = 0.1 \alpha_{-1}$, $\beta = 0.1 \beta_{-1}$,
$S \equiv 3 \pi \Sigma(a_{\rm d}) \nu(a_{\rm d})/{\dot M}_{\rm Edd}$ and 
$\theta = 0.2 \theta_{0.2}$. Here ${\dot M}_{\rm Edd} = 4 \pi M m_p/(\eta 
\sigma_T)$ is the Eddington
accretion rate, $\sigma_T$ is the Thomson cross-section for electron 
scattering, $\eta$ is the radiative efficiency and $\theta < 1$ is a porosity 
correction factor applied to the 
scattering-dominated optical depth ~\cite{Tur04}. The 
radiative efficiency  
$\eta \equiv L_{\rm {Edd}}/{\dot M}_{\rm Edd}$ may be set to 
1/12 for a Newtonian thin disk disk with an innermost stable
circular orbit (ISCO) at  
$r_{\rm isco} = 6 M$, the value appropriate for a nonspinning 
black hole.

Most earlier work has focused on the late binary epoch {\it following} 
decoupling when $a < a_d$. During this epoch the outer disk remains 
almost frozen until after binary inspiral and coalescence, at which point 
the disk diffuses inward
and fills up the hollow (in the case of equal-mass black holes), 
or gap (in the case of a low-mass companion), on a (slow) viscous 
timescale (see, e.g.,~\cite{TanM10,Sha10} and references therein). 
Eventually the gas 
reaches the  ISCO of the remnant black hole, where all torques and surface 
densities vanish in a first approximation. Ultimately the disk settles
into steady-state equilibrium around the black hole remnant.
In this paper we shall be concerned primarily in the early binary epoch
{\it prior} to decoupling when $a > a_d$. As we have described, 
during this epoch it is the binary orbit
that remains nearly frozen while 
the disk adjusts to the combined tidal and viscous torques, 
the effects of which balance each other in steady state. 
Thus, up until the BHBH orbit reaches decoupling, the disk 
evolves quasistatically as the orbit shrinks.

Because $t_{\rm orb}(r) \ll t_{\rm vis}(r)$ over the entire history of a 
quasistationary disk, the disk radial velocity satisfies $v_r \ll r \Omega_K(r)$ and 
the full hydrodynamic equations reduce to secular 
(conservation) equations to describe the orbit-averaged evolution of a Keplerian thin 
disk~\cite{Pri81}.  While obtaining the spiral 
density wave perturbations induced by the tidal torques 
near the edge of the disk does require the full set of 
hydrodynamical equations, deducing the underlying orbit-averaged disk profile 
does not. For the epoch prior to decoupling these secular 
equations can be solved in
steady state to give the quasisteady disk structure 
for each value of the separation $a > a_d$. We perform this calculation
below in the low-mass limit where $q < 1$.

\subsection{Key Equations}
\subsubsection{Disk Evolution}
The evolution of a geometrically thin, nearly Keplerian disk is determined by 
combining the equation of mass conservation,
\begin{equation}
\frac{\partial \Sigma}{\partial t} + \frac{1}{r} 
\frac{\partial (r \Sigma v_r)}{\partial r} = 0
\end{equation}
with the equation of angular momentum conservation,
\begin{equation}
\frac{\partial (\Sigma r^2 \Omega)}{\partial t} + \frac{1}{r}
\frac{\partial (r \Sigma v_r r^2 \Omega)}{\partial r} = \frac{1}{2 \pi r}
\frac{\partial G}{\partial r}
\end{equation}
to obtain an evolution equation for the surface density,
$\Sigma(t,r)$, 
\begin{equation}
\label{eq:sigevol1}
\frac{\partial \Sigma}{\partial t} = - \frac{1}{2 \pi r} 
\frac{\partial}{\partial r} 
\left[
\left( \frac{\partial (r^2 \Omega)}{\partial r} \right)^{-1}
\frac{\partial G}{\partial r} 
\right]
\ .
\end{equation}
Here $G \equiv -T_{\rm vis} + T_{\rm tid}$ is the total torque, $T_{\rm vis}$ is the
viscous torque, $T_{\rm tid}$ is the tidal torque on the disk from the presence of 
the secondary, 
and $\Omega = \Omega_K$ is the orbital frequency. The viscous torque density is
given by the standard equation~\cite{Pri81,FraKR02,ArmN02,ChaSMQ10}   
\begin{equation}
\frac{\partial T_{\rm vis}}{\partial r} = 
-\frac{\partial}{\partial r} \left( 2 \pi r^3 \nu \Sigma 
\frac{\partial \Omega}{\partial r} \right).
\label{eq:dr_Tvis}
\end{equation}
We approximate the (orbit-averaged) tidal torque density 
by using the expression adopted by Armitage and Natarajan~\cite{ArmN02}   
\beq
\label{eq:AN}
\frac{\partial T_{\rm tid}}{\partial r}  = 2 \pi \Lambda  \Sigma r \,
\eeq 
where $\Lambda(r,a)$ is given by
\beq
\label{eq:torqtid}
\Lambda = \left \{ \begin{array}{ll}
-\left (f q^2 M/2r \right) \left( r /\Delta_p \right)^4,   & r < a \\
+\left (f q^2 M/2r \right) \left( a /\Delta_p \right)^4,  & r > a \\
     \end{array} \right. \ .
\eeq

In Eq.~(\ref{eq:torqtid}) $f$ is a dimensionless 
normalization factor 
and $\Delta_p$ is given by $\Delta_p = {\rm max}(|r-a|, h)$.
Calibrating the above expression for the tidal field 
against high-resolution, hydrodynamical  
simulations in two-dimensions for a low-mass, black hole secondary 
interacting with an outer 
accretion disk, Armitage and Natarajan find that the 
value $f \approx 0.01$ best fits the 
simulation results. Equations~(\ref{eq:AN}) and (\ref{eq:torqtid})
furnish a reasonable analytic approximation 
to the results obtained from summing over the pointlike 
contributions from the Lindblad resonances in the disk~\cite{GolT80,HouW84}. 
(Similar, but slightly different, forms for the tidal 
torque also have been used in the literature; 
see, e.g.,~\cite{LinP86,War97,Cha08,ChaSMQ10}. For an analysis in general 
relativity, see~\cite{Hirata10a,Hirata10b}.)
Assembling the above expressions then yields the final evolution equation 
\beq
\label{eq:sigevol2}
\frac{\partial \Sigma}{\partial t} = \frac{1}{r} \frac{\partial}{\partial r}
\left[ 
3 r^{1/2} \frac{\partial}{\partial r} \left( r^{1/2} \nu \Sigma \right)
- \frac{2 \Lambda \Sigma r^{3/2}}{M^{1/2}} 
\right] \ .
\eeq 

The rate at which the secondary black hole migrates is determined both
by back-reaction to the tidal torquing of the disk and by gravitational wave 
emission, 
\beq
\label{eq:aevol}
\dot a = \dot a_{\rm tid} + \dot a_{\rm GW}\ ,
\eeq
where
\beq
\label{eq:atidevol}
\dot a_{\rm tid} = -\frac{4 \pi a^{1/2}}{M^{3/2}q} 
\int_{r_{\rm isco}}^{r_{\rm out}} r \Lambda \Sigma dr\ ,
\eeq
and where $a_{\rm GW}$ is given by Eq.~(\ref{eq:GW}). In 
Eq.~(\ref{eq:atidevol}) the integration is over the entire disk, although
most of the contribution from tidal torques 
arises close to the gap boundaries near $ r \approx a$.

The accretion rate onto the primary may be calculated from
\begin{eqnarray}
\label{eq:mdot}
\dot M (t,r) &=& 2 \pi r \Sigma (-v_r), \cr
       &=& - \left[ \frac {(r^2 \Omega)}{\partial r} \right]^{-1} 
\frac{\partial G} {\partial r}\ .
\end{eqnarray} 
Combining Eqs.~(\ref{eq:sigevol1}) and ~(\ref{eq:mdot}) yields
\begin{equation}
\label{eq:mdot2}
\frac{\partial \Sigma}{\partial t} 
=  \frac{1}{2 \pi r}
\frac{\partial}{\partial r} \dot M\ .
\end{equation}

The coupled evolution Eqs.~(\ref{eq:sigevol2}) and  (\ref{eq:aevol}) have 
been integrated in time previously to explore select cases 
(see e.g.~\cite{ArmN02,ChaSMQ10}). 
Here we want to solve them for quasistationary flow in general. 
We will thereby obtain initial data for the construction of
disks around low-mass binaries prior to decoupling that is valid in 
general cases.
Thus we set $ a = {\rm constant}$ and  $\partial \Sigma / \partial t = 0$  
in Eq.~(\ref{eq:sigevol2}) to 
obtain the density profile.  According to Eq.~(\ref{eq:mdot2}) the
resulting quasisteady accretion rate $\dot M$ is independent of $r$ 
and can be obtained by solving Eq.~(\ref{eq:mdot}) once the density profile 
has been determined.

In steady state, Eq.~(\ref{eq:sigevol2})
becomes a second-order elliptic equation in $r$, for which we 
impose the following boundary conditions:
\begin{equation}
\label{eq:bc}
\mbox{b.c.'s}: \ \ \ \nu \Sigma = \left \{ \begin{array}{ll}
(\nu \Sigma)_{\rm out},&
r = r_{\rm out}\\
 0,& r = r_{\rm isco}
\end{array} \right. .
\end{equation}
In Eq.~({\ref{eq:bc}) $r_{\rm out}$ is the outer radius
of the disk.  
We set the ISCO in the disk equal to
$r_{\rm isco} = 6M$,
the value appropriate for a nonspinning black hole 
remnant.
Typically, $r_{\rm out} \gg r_{\rm isco}$ and in some cases we shall take 
$r_{\rm out} \rightarrow \infty$.

As mentioned above, the
accretion rate $\dot M$ in steady state is independent of $r$.
In steady state Eq.~({\ref{eq:sigevol2}) admits a first integral which,
when combined with Eq.~(\ref{eq:mdot}) yields the first-order 
ODE
\begin{equation}
\label{eq:mdot3}
\dot M = 2 \pi \left[ 
3 r^{1/2}\frac{d \left( r^{1/2} \nu \Sigma \right)} {d r} 
- \frac{2 \Lambda \Sigma r^{3/2}}{M^{1/2}} 
\right ] 
= {\rm constant}
\end{equation}
While we could solve the second-order Eq.~(\ref{eq:sigevol2}) 
in steady state for the density profile, followed
by Eq.~(\ref{eq:mdot}) for the accretion rate, it is simpler 
to integrate the first-order Eq.~(\ref{eq:mdot3}). We do this 
in Sec.~\ref{subsec:gensoln} below.

\subsubsection{Electromagnetic Radiation}
\label{sec:em1}

The local radiated emission from the disk arises both from viscous and tidal
dissipation. The rate of viscous dissipation per unit surface area is
~\cite{Pri81}
\begin{equation}
\label{diss}
D_{\rm vis}(t,r)=\frac{9}{8}\nu \Sigma \frac{M}{r^3}.
\end{equation}
The integrated rate of tidal dissipation may be obtained from the 
change in the binary binding energy due to tidal forces, 
\begin{equation}
\dot E_{\rm tid}(t) = 
 \frac{M^2 q}{2a} \left(\frac{|\dot a|}{a}\right)_{\rm tid}\ ,
\label{eq:Edottid}
\end{equation}
where $(\dot a/a)_{\rm tid}$ is given by Eq.~({\ref{eq:atidevol}). 
We follow ~\cite{ChaSMQ10} and assume that tidal dissipation results from
the local damping of spiral density waves that mediate the binary-disk 
interaction and is thus proportional to the local generation of these waves,
\begin{equation}
D_{\rm tid}(t,r) = \frac{1}{4 \pi r}\dot E_{\rm tid} 
\frac{|dT_{\rm tid}/dr|}{\int_{r_{\rm isco}}^{r_{\rm out}} dr |dT_{\rm tid}/dr|}\ .
\label{eq:Dtid}
\end{equation}

Approximating the 
emission as thermal blackbody radiation,
the local disk surface temperature $T_s(t,r)$ may be equated to the 
effective temperature, which is determined from the total
dissipation rate $D = D_{\rm vis} + D_{\rm tid}$ according to
\begin{equation}
\label{Tsurf}
\sigma T_s^4(t,r)=D(t,r)
\end{equation}
where $\sigma$ is the Stefan-Boltzmann constant ~\cite{COLOR}. 

Given the surface temperature, the quasistationary specific flux $F_{\nu}(t)$ 
measured by an observer at 
distance $d$ whose line of sight makes an angle $i$ to the normal to the disk 
plane is determined by integrating over the entire disk surface, 
\begin{equation}
\label{flux1}
F_{\nu}(t) = \frac{2 \pi ~{\rm cos}~i} {d^2} 
\int_{r_{\rm isco}}^{\infty} B_{\nu}(T_s(t^{\prime},r)) r dr, 
\end{equation}
where  $B_{\nu}(T_s(t^{\prime},r))$ is the Planck function, $t^{\prime} 
= t-d$ is retarded time and $\nu$ is the photon frequency~\cite{FREQ}.
Equation~(\ref{flux1})
is best evaluated in terms of a nondimensional function 
$f^*(t^{\prime},x)$ of nondimensional frequency $x$ according to
~\cite{Sha10}
\begin{equation}
\label{flux}
F_{\nu}(t) =  \frac{2 \pi ~{\rm cos}~i}{d^2} \frac{15}{\pi^5} \frac{\sigma T_*^4}
{\nu_*} r_{\rm isco}^2 f^*(t^{\prime},x),
\end{equation}
where
\begin{equation}
\label{f*}
f^*(t^{\prime},x) \equiv \int_1^{\infty} du ~u \frac{x^3}
{{\rm exp} \left(x T_*/T_s \right) -1},
\end{equation}
and where we have introduced the parameters
\begin{eqnarray}
\label{T*}
\sigma T_*^4 &\equiv& 3 M \dot{M}_{\rm rem}/8 \pi r_{\rm isco}^3, 
\ \ \ h \nu_* \equiv kT_*, \cr
x &\equiv& h \nu /k T_* = \nu/\nu_*, \ \ \ u \equiv r/r_{\rm isco}.
\end{eqnarray}
Here $\dot{M}_{\rm rem}=3 \pi (\nu \Sigma)_{\rm out}$ is the 
stationary accretion rate of an infinite disk onto the remnant black hole
following merger [see Eq.~(\ref{eq:sigfinal})].
The quantity $T_*$ provides a convenient estimate of the characteristic 
temperature in the main radiating region near $r_{\rm isco}$ of the final 
equilibrium disk following merger, and $h \nu_*$ is the 
characteristic frequency of the emitted thermal radiation from this region.  
The specific luminosity $L_{\nu} (t)$, summing over both surfaces of the disk,
is related to $F_{\nu}(t)$ according to
\begin{equation}
L_{\nu} (t) = \frac{2 \pi d^2}{{\rm cos}~i} F_{\nu}(t).
\end{equation}
The total luminosity $L(t)$ integrated over all frequencies is then given by
\begin{equation}
\label{eq:lum1}
L(t) = \int_0^{\infty} d \nu L_{\nu}(t) = \frac{60}{\pi^3} \sigma T_*^4 
r_{\rm isco}^2 \int_0^{\infty} f^*(t^{\prime},x) dx
\end{equation}

Given the quasistationary density profile, we wish to perform these quadratures
for the quasistationary electromagnetic spectrum prior to
decoupling.  Equation~(\ref{eq:lum1}) yields 
\begin{equation}
\label{eq:lum2}
L = 2 \int_ {r_{\rm isco}}^{r_{\rm out}} D_{\rm vis} 2 \pi r dr  
+ \dot E_{\rm tid}\ .
\end{equation}

We note that  
the final stationary equilibrium disk following merger yields well-known 
analytic density and 
temperature profiles, as well as analytic integrated fluxes and 
luminosities (see, e.g., \cite{Sha10}),
and these quantities provide useful checks on the numerical quadratures.
For example, after the merger, the tidal torque is absent and 
the equilibrium density profile is given by the familiar 
Shakura-Sunyaev result~(\cite{ShaS73}; see also ~\cite{NovT73,Pri81,FraKR02} 
and references therein) for a thin disk around a single black hole remnant,
\begin{eqnarray}
\label{eq:sigfinal}
\nu \Sigma &=& \left( \nu \Sigma \right)_{\rm out} 
\frac{\left( 1- r_{\rm isco}^{1/2}/ r^{1/2} \right )}
{\left( 1- r_{\rm isco}^{1/2}/ r_{\rm out}^{1/2} \right )}, 
\ \ \  [{\rm post-merger}] \cr
          &=& \frac{\dot M}{3 \pi} 
\left( 1 - r_{\rm isco}^{1/2}/r^{1/2} \right). 
\end{eqnarray}
The corresponding luminosity is given by 
\begin{eqnarray}
\label{eq:Lfinal}
L &=& \frac{\dot M M}{2 r_{\rm isco}} 
\left ( 1 - 3 \frac{r_{\rm isco}}{r_{\rm out}}
+2 \frac{r_{\rm isco}^{3/2}}{r_{\rm out}^{3/2}} \right), 
\ \ \  [{\rm post-merger}] \cr
  &=& \frac{\dot M M}{2 r_{\rm isco}},  \ \ \ 
r_{\rm out} \rightarrow \infty.
\end{eqnarray}

\subsubsection{Structure Equation: Nondimensional Units and Scaling}
\label{sec:nondim}

To solve Eq.~(\ref{eq:sigevol2}) in steady state for general cases, 
and to help identify scaling behavior, it is convenient to introduce 
the following nondimensional variables:
\begin{eqnarray}
\label{eq:nondim}
s &=& (r/r_{\rm out})^{1/2}, \ \ s_1 = (a/r_{\rm out})^{1/2}, 
\ \ s_2 = (r_{\rm isco}/r_{\rm out})^{1/2}, \cr 
\ \ \bar {\Sigma} &=& \Sigma/\Sigma_{\rm out}, 
\ \ \bar{\nu}=\nu/\nu_{\rm out}, \ \ y = s \bar{\Sigma},  
\ \ \ybar = \nubar y, \cr
\ \ \ \bar {h} &=& h/r, \ \ \tau=t/2 t_{\rm vis}(r_{\rm out}), 
\ \ {\dot m}  = {\dot M}/(3 \pi \nu_{\rm out} \Sigma_{\rm out}).
\end{eqnarray}
In terms of these variables, equation~(\ref{eq:sigevol2}) becomes
\begin{equation}
\label{eq:ndsigevol2}
\frac{\partial y}{\partial \tau} = \frac{1}{s^2} 
\frac{\partial^2 (\bar {\nu} y)} {\partial s^2} 
- \frac{1}{s^2} \frac{\partial}{\partial s} 
\left \{
g^*(s) y \left[ \frac{1}{{\rm max} (|s^2 - s_1^2|, s^2 \bar {h})} \right]^4 
\right \}, 
\end{equation}
where
\beq
\label{eq:g}
g=\frac{2}{3}\frac{f q^2 M^{1/2} r_{\rm out}^{1/2}}{\nu_{\rm out}}
\eeq
and where
\beq
\label{eq:g*}
  g^*(s) = \left \{ \begin{array}{ll}
        g s_1^8      & s>s_1 \\
        -g s^8     & s<s_1
        \end{array} \right. \ .
\eeq 

Evaluating  Eq.~(\ref{eq:ndsigevol2}) in steady state yields
the second-order ODE
\beq  
\label{eq:ODE2}
\frac{d^2}{ds^2}(\nubar y) - \frac{d}{ds} \left \{ g^*(s)
 \left[ \frac{1}{\max(|s^2-s_1^2|,s^2 \bar{h} )}\right]^4 y \right \} = 0 \ ,
\eeq 
which must be solved for $s\in [s_2,1]$ subject the boundary conditions
\begin{equation}
\label{eq:bc2}
\mbox{b.c.'s}: \ \ \ y = \ybar = \left \{ \begin{array}{ll}
1,&
s = 1\\
 0,& s = s_2
\end{array} \right. .
\end{equation}
Equation~(\ref{eq:ODE2}) admits a first integral, just as in 
Eq.~({\ref{eq:mdot}), so that we need only solve the first-order ODE \beq
  \frac{d \ybar}{ds} - f(s) \ybar = \dot{m} \ ,
\label{eq:disk-in-out}
\eeq
where
\beq
 f(s) = 
\frac{g^*(s)}{\nubar(s)} \left[ 
\frac{1}{\max(|s^2-s_1^2|,s^2 \bar {h})}\right]^4 \ .
\label{eq:newfs}
\eeq

\section{Structure Equation: Steady-State Solution}

\subsection{General Solution}
\label{subsec:gensoln}

To solve Eq.~(\ref{eq:disk-in-out}) we introduce the function 
\beq
  F(s) = \int_s^1 f(s') ds' .
\label{def:Fs}
\eeq
The ODE can be rewritten as 
\beq
  \frac{d}{ds} \left( e^F \ybar \right) = \dot{m} e^{F} ,
\eeq
and is readily integrated to give 
\beq
  \ybar(s) = e^{-F(s)} \left[ b - \dot{m} \int_s^1 e^{F(s')} ds'\right] \ .
\eeq
The boundary condition  $\ybar(1)=1$ gives $b=1$, while the condition
 $\ybar(s_2)=0$ gives
\beq
  \dot{m} = \left[ \int_{s_2}^1 e^{F(s)} ds\right]^{-1} \ .
\label{eq:mdot-in-out-new}
\eeq
Hence the solution is 
\beq
  y(s) = \frac{e^{-F(s)}}{\nubar(s)} \left[ 1 - \dot{m} \int_s^1 e^{F(s')} ds' 
\right] \ .
\label{eq:ysol-in-out-new}
\eeq

To compute $\dot{m}$ and $y(s)$ numerically, it is convenient to introduce the 
function 
\beq
  G(s) = \int_s^1 e^{F(s')} ds' \ .
\label{eq:Gs}
\eeq
Functions $F$ and $G$ satisfy the coupled ODEs
\beqn
  F'(s) &=& -f(s) , \label{eq:Fprime-new}  \\
  G'(s) &=& -e^{F(s)} ,  \label{eq:Gprime-new}
\eeqn
with the initial conditions $F(1)=G(1)=0$. Equations~(\ref{eq:mdot-in-out-new}) 
and (\ref{eq:ysol-in-out-new}) can be written as 
\beqn
  \dot{m} &=& \frac{1}{G(s_2)} \ , \label{eq:mdotG2} \\ 
  y(s) &=& \frac{e^{-F(s)}}{\nubar(s)} \left[ 1 - \frac{G(s)}{G(s_2)}\right] \ .
\label{eq:yanalytic}
\eeqn

In some cases, $G(s)$ may become large when $s$ approaches $s_1$. 
To treat this complication we introduce the function $H(s)=\ln[ G(s)]$, which 
satisfies the ODE 
\beq
  H'(s) = \frac{G'(s)}{G(s)} = -e^{F(s)-H(s)} \ .
\label{eq:Hprime-new}
\eeq
In practice, the coupled ODEs (\ref{eq:Fprime-new}) and (\ref{eq:Gprime-new}) are 
integrated from $s=1$ to $s=s^*$, where $s_1 < s^* < 1$. Then Eqs.~(\ref{eq:Fprime-new}) 
and (\ref{eq:Hprime-new}) are integrated from $s=s^*$ to $s=s_2$ with the 
initial condition $H(s^*)=\ln[G(s^*)]$. Having computed the functions 
$F$ and $H$ for $s\in [s_2,1]$, $\dot{m}$ and $y(s)$ are obtained by 
\beqn
\label{eq:struceqn1}
  \dot{m} &=& e^{-H(s_2)} \ , \\ 
\label{eq:struceqn2}
  y(s) &=& \frac{e^{-F(s)}}{\nubar(s)} \left[ 1 - e^{H(s)-H(s_2)} \right]  \ .
\eeqn

We shall study the solution by numerically 
integrating the above equations for different
choices of parameters in Section~\ref{sec:numinteg} below.  

\subsection{Limiting Cases and Asymptotic Behavior}

Before obtaining numerical solutions for general cases 
it is instructive to
evaluate Eq.~(\ref{eq:disk-in-out}) analytically for the disk structure 
in limiting regimes. We first observe that the parameter $g$ defined in
Eq.(\ref{eq:g}) may be evaluated as
\beq
\label{eq:g2}
g = 2 \pi f q^2 \left( \frac{t_{\rm vis}}{t_{\rm orb}} \right)_{\rm out}
\eeq
where we have used Eqs.~(\ref{eq:torb}) and (\ref{eq:tvis}). 
We also find it useful to introduce another nondimensional parameter 
\beq
  \tilde{g} = \frac{2 \pi f q^2}{2\bar{h}^3} \left( \frac{t_{\rm vis}}{t_{\rm orb}} \right)_{r=a} 
 = \frac{fq^2 M^{1/2} a^{1/2}}{3 \nu(a) (h/a)^3} .
\label{eq:gtilde}
\eeq
The physical meaning of $\tilde{g}$ can be understood as follows. 
Consider a ring of disk material of width $h$ near $r=a$. The tidal 
torque action on the ring is [see Eq.~(\ref{eq:AN})] 
$T_{\rm tid}(a) \approx 2\pi \Lambda \Sigma a h$. The viscous torque is 
$T_{\rm vis}(a) = -2\pi a^3 \nu(a) \Sigma  \partial_r \Omega_K(a)$ 
[see Eq.~(\ref{eq:dr_Tvis})]. 
Hence 
\beq
  \frac{T_{\rm tid}(a)}{T_{\rm vis}(a)} \sim \frac{2h|\Lambda(a)|}{3\nu(a)} \left( 
\frac{a}{M}\right)^{1/2} = \tilde{g} .
\eeq
The parameter $\tilde{g}$ therefore measure 
the relative importance of the tidal to viscous torques at 
$r=a$: when $\tilde{g}$ is small 
the tidal torque is unimportant, otherwise it plays a significant role. 
The variables $f$, $q$ and $\bar{h}$ in Eq.~(\ref{eq:gtilde}) are each always 
less than unity, while the ratio $t_{\rm vis}/t_{\rm orb}$ is always 
larger than unity; hence $\tilde{g}$ can vary between 
zero and infinity, depending on the model (the same conclusion also applies 
to the parameter $g$). 

\subsubsection{Negligible Tidal Torques}
\label{sec:Notidal}

In the limiting regime $\tilde{g} \rightarrow 0$ tidal torques are negligible and
Eq.~(\ref{eq:disk-in-out}), may be solved together with boundary 
conditions (\ref{eq:bc2}) to yield
\beq
\label{eq:ynotorque}
\bar y = \dot{m} (s -s_2), \ \  (g=0),
\eeq
and 
\beq
\label{eq:mdotnotorque}
\dot{m} = \frac{1}{1 - s_2}\ .
\eeq
Restoring units, Eqs.~(\ref{eq:ynotorque}) and (\ref{eq:mdotnotorque}) 
translate to 
\beq
\label{eq:signotorque}
\nu \Sigma = \frac{\dot M}{3 \pi} 
\left[ 1 - r_{\rm isco}^{1/2}/r^{1/2} \right]
\eeq
and
\begin{eqnarray}
\label{eq:mdotnotorque2}
\frac{\dot M}{3 \pi (\nu \Sigma)_{\rm out}} &=& 
\frac{1}{1- r_{\rm isco}^{1/2}/r_{\rm out}^{1/2}} \cr  
&=& 1, \ \ \ r_{\rm out} \rightarrow \infty,
\end{eqnarray}
respectively. As expected, when tidal torques arising from the companion 
are negligible, Eqs.~(\ref{eq:signotorque}) and (\ref{eq:mdotnotorque2}) 
reduce to Eq.~(\ref{eq:sigfinal}), the 
quasistationary profile and accretion rate for a disk around a
single black hole. 

\subsubsection{Strong Tidal Torques}
\label{sec:strongtid}

In another limiting regime $\tilde{g} \gg 1$, tidal torques are strong and 
we will see in the following analysis that the accretion is shut off. 

In the thin-disk limit $\bar{h} \rightarrow 0$, the function $f(s)$ changes 
from a large negative value to a large positive value at $s=s_1$. The function 
$F(s)$ defined in Eq.~(\ref{def:Fs}) thus has a narrow peak near $s=s_1$. 
Hence we may approximate $\dot{m}$ 
in Eq.~(\ref{eq:mdot-in-out-new}) by integrating $e^F$ over the small region 
around $s=s_1$ where $|f(s)|$ reaches maximum. We first approximate $f(s)$ near $s=s_1$ by a step function
\beq
  f(s) \approx \left \{ \begin{array}{ll} 
 	\frac{2\tilde{g}}{s_1 \bar{h}} , & s_1 < s < s^+ , \\ 
	-\frac{2\tilde{g}}{s_1 \bar{h}} , & s^- < s <s_1 , \\ 
  \end{array} \right. ,
\eeq
where $s^+=s_1/\sqrt{1-\bar{h}} \approx s_1(1+\bar{h}/2)$ and 
$s^- = s_1/\sqrt{1+\bar{h}} \approx s_1(1-\bar{h}/2)$. 
Using Eq.~(\ref{def:Fs}) we find 
\beq
  F(s) \approx F_1 + \tilde{g} \left( 1-\left|\frac{2(s-s_1)}{s_1 \bar{h}}\right| \right) 
\ \ \ \mbox{for } s^- < s < s^+ ,
\label{eq:Fslgt}
\eeq
where 
\beqn
  F_1 &=& \int_{s^+}^1 f(s) ds \cr
      &\sim & \frac{gs_1^8}{\bar{\nu}(s_1)} \int_{s^+}^1 \frac{ds}{(s^2-s_1^2)^4} \cr
      &=& \frac{\tilde{g}}{3} [ 1 + O(\bar{h})] .
\label{eq:F1lgt}
\eeqn
The accretion rate $\dot{m}$ can then be approximated by 
\beqn
\frac{1}{\dot{m}} &\sim & \int_{s^-}^{s^+} e^{F(s)} ds , \cr 
 &\sim & e^{\tilde{g}/3} \int_{s^-}^{s^+} ds\, \exp \left[ \tilde{g}
\left( 1 -2\left|\frac{s-s_1}{s_1 \bar{h}} \right|\right) \right] , \cr
 &=& \frac{s_1 \bar{h}}{\tilde{g}} 
e^{\tilde{g}/3} \left( e^{\tilde{g}} -1 \right) .
\eeqn
For $e^{\tilde{g}} \gg 1$, we have 
\beq
  \dot{m} \sim \frac{\tilde{g}}{s_1 \bar{h}} e^{-4\tilde{g}/3} .
\label{eq:mdothg}
\eeq
Hence when the tidal torques are strong, the accretion is shut off. 
Our numerical calculation confirms that Eq.~(\ref{eq:mdothg}) is accurate 
when $e^{\tilde{g}} \gg 1$ and $\bar{h} \ll 1$. 

In Sec.~\ref{sec:Notidal}, we see that $\dot{m} \approx 1$ for weak 
tidal torques ($\tilde{g} \ll 1$). The above analysis shows that 
$\dot{m} \approx 0$ 
for strong tidal torques ($e^{\tilde{g}}\gg 1$). The critical transition between 
these two regimes occurs at $\tilde{g}=\tilde{g}_c$, which can be 
estimated by the condition 
\beq 
\frac{\tilde{g}_c}{s_1 \bar{h}} e^{-4\tilde{g_c}/3} \sim 1 .
\label{cond:gtc}
\eeq
For $a=100M$, $r_{\rm out}=10^5M$ ($s_1=10^{-3/2}$) and $\bar{h}=0.1$, Eq.~(\ref{cond:gtc}) gives 
$\tilde{g}_c \sim 6$. We find in Sec.~\ref{sec:numinteg} below that 
$\tilde{g}_c$ lies between 5 and 10 for a power law viscosity $\nu(r)\propto r^n$, which is 
consistent with this simple analysis. 

The density profile in the strong tidal torque regime is very different 
from that in the weak tidal torque regime. We will see in 
Sec.~\ref{sec:asysigma} below that in the asymptotic region in which 
$r\gg a$, the density profile is given by 
\beq
  \Sigma(r) \approx \Sigma_{\rm out} \frac{\nu_{\rm out}}{\nu(r)} 
\left( \frac{r}{r_{\rm out}} \right)^{-1/2} = \Sigma_{\rm out}\left( \frac{r}{r_{\rm out}} \right)^{-(n+1/2)}
\eeq
for a power law viscosity. To estimate $\Sigma$ near $r=a$, we first note from 
Eq.~(\ref{eq:Gprime-new}) that $G(s)<G(s_2)$ for $s>s_2$ ($r>r_{\rm ISCO}$). 
Hence $y(s) < e^{-F(s)}/\nubar(s)$. For $s^-<s<s^+$, Eqs.~(\ref{eq:Fslgt}) 
and (\ref{eq:F1lgt}) give 
\beq
  \frac{\Sigma(r)}{\Sigma_{\rm out}} 
< \frac{1}{s \nubar(s)}
\exp \left\{ -\tilde{g} \left[ \frac{4}{3} - \left| \frac{2(s-s_1)}{s_1\bar{h}} 
\right| \right] \right \}  \ll 1
\eeq
for large $\tilde{g}$. For $s<s^-$, we combine Eqs.~(\ref{eq:Gs}), (\ref{eq:mdotG2}) 
and (\ref{eq:yanalytic}) to write 
\beq
  y(s) = \frac{\dot{m}}{\nubar(s)} \int_{s_2}^s e^{F(s')-F(s)} ds' .
\eeq
It follows from Eqs.~(\ref{eq:Fprime-new}), (\ref{eq:newfs}) and (\ref{eq:g*}) that 
$F(s')-F(s)<0$ for $s'<s<s_1$. Hence 
\beq
  \frac{\Sigma(r)}{\Sigma_{\rm out}}  < \dot{m}\left( \frac{r}{r_{\rm out}}\right)^{-n} 
\left( 1 - \sqrt{\frac{r_{\rm isco}}{r}} \right) \ll 1 \ \ \ (r<a)
\eeq
for small $\dot{m}$ (large $\tilde{g}$). 

We therefore see that the density plummets for $r \lesssim a$ when the tidal torques 
are strong. This behavior can be understood as follows. The tidal torque $T_{\rm tid}$ pushes 
the disk matter radially outwards for $r>a$ and inwards for $r<a$; whereas the viscous 
torque $T_{\rm vis}$ always drags the matter inwards. The tidal torques are strongest 
near $r=a$, which tend to suppress the inflow and create a gap there.
When $T_{\rm tid}$ is sufficiently strong, the inflow is strongly suppressed 
near $r=a$. Inside the binary's orbit ($r<a$), both $T_{\rm tid}$ and $T_{\rm vis}$ push matter 
inwards, so that when $T_{\rm tid}$ is strong matter cannot accumulate in the inner region. Hence in steady 
state, there is practically no inner disk and the density of the outer disk near $r=a$ 
is also very small. 

\subsubsection{Asymptotic Behavior}
\label{sec:asysigma}

As shown above, when tidal torques are present, the accretion rate 
may be much lower
than the value quoted in Eqs.~(\ref{eq:mdotnotorque}) and 
(\ref{eq:mdotnotorque2}) for accretion onto a single black hole. The rate 
$\dot m$ depends sensitively on $\tilde{g}$ and requires the full numerical
solution to determine accurately. But in the asymptotic region $r \gg a$ it is always 
possible to express the disk structure in terms of $\dot m$ even 
when tidal torques are present, as we now show.   

Whenever $1 \geq s \gg s_1$, Eqs.~(\ref{eq:disk-in-out}) 
and (\ref{eq:newfs}) reduce to 
\beq
\label{eq:sggs1}
  \frac{d \ybar}{ds} - \frac{g}{\nubar} \left( \frac{s_1}{s} \right)^8 \ybar = \dot{m} \ .
\eeq
Consider the asymptotic regime where 
\beq
\label{eq:sasym}
\frac{g s}{\bar \nu} \left( \frac{s_1}{s} \right)^8 \ll 1,
\eeq
in which case the second term in Eq.~(\ref{eq:sggs1}) is typically 
much less than the first term and therefore can be dropped.
The solution to the resulting equation is
\beq
\label{eq:asym}
\bar y \approx \dot m s + 1 - \dot m,
\eeq
or 
\beq
\label{eq:asym2}
\nu \Sigma \approx \left( \nu \Sigma \right)_{\rm out} 
\left[ \dot m + \left(1-\dot m \right) 
\left( \frac{r}{r_{\rm out}} \right)^{-1/2}  \right],
\eeq
where we have used the boundary condition $\bar y(1) = 1$ (Eq.~\ref{eq:bc2}).
As we have argued in Sec.~\ref{sec:strongtid}}, 
whenever the tidal torque parameter 
satisfies $\tilde{g} \lesssim \tilde{g}_c$ we have $\dot m \approx 1$, while 
whenever $\tilde{g} \gtrsim \tilde{g}_c$ we have $\dot m \ll 1$. Not surprisingly, 
whenever tidal torques are unimportant, the asymptotic density profile
reflected by Eqs.~(\ref{eq:asym}) and ~(\ref{eq:asym2}) is 
the same  as the asymptotic profile given by 
Eqs.~(\ref{eq:ynotorque}) and ~(\ref{eq:signotorque})  
(i.e., $\nu \Sigma \approx constant \approx (\nu \Sigma)_{\rm out} 
\approx \dot M/3 \pi$), but has a  very different fall-off otherwise.

As an application, take the disk viscosity to be a power-law profile,
\beq
\label{eq:vispow}
\nu(r) \propto r^n, \ \ \ n \neq -7/2,
\eeq
or $\nubar(s) = s^{2n}$ with $n\neq -7/2$. Then Eq.~(\ref{eq:sggs1}) 
becomes
\beq
  \frac{d}{ds} \left[ q(s) \ybar \right] = q(s) \dot{m} \ ,
\label{eq:asyeq}
\eeq
where 
\beq
  q(s) = \exp \left[ \frac{g}{(2n+7) s^{2n-1}} 
\left( \frac{s_1}{s}\right)^8\right] . \\ 
\eeq
Consider the asymptotic region where 
\beq
  \frac{g}{(2n+7) s^{2n-1}} \left( \frac{s_1}{s}\right)^8 \ll 1 ,
\eeq
or 
\beq
  s \gg \left( \frac{g s_1^{2n-1}}{2n+7} \right)^{1/(2n+7)} s_1 \ .
\label{eq:s_asy}
\eeq
For example, if $n=1/2$, our fiducial case below, Eq.~(\ref{eq:s_asy}) 
reduces to 
\beq
  s \gg \left( \frac{g}{8}\right)^{1/8} s_1 \ .
\eeq
In the asymptotic region defined by Eq.~(\ref{eq:s_asy}), 
$q(s)\approx 1$ and Eq.~(\ref{eq:asyeq}) again can be integrated to give 
Eq.~({\ref{eq:asym}), which now yields
\beq
\frac{\Sigma}{\Sigma_{\rm out}} = \frac{\bar y(s)}{s^{2n+1}} \approx \dot{m} 
\left( \frac{r}{r_{\rm out}}\right)^{-n} 
+\ (1-\dot{m}) \left( \frac{r}{r_{\rm out}}\right)^{-(n+1/2)} \ .
\label{eq:sigasym}
\eeq

\subsubsection{Numerical Integrations}
\label{sec:numinteg}

\begin{figure}
\includegraphics[width=9cm]{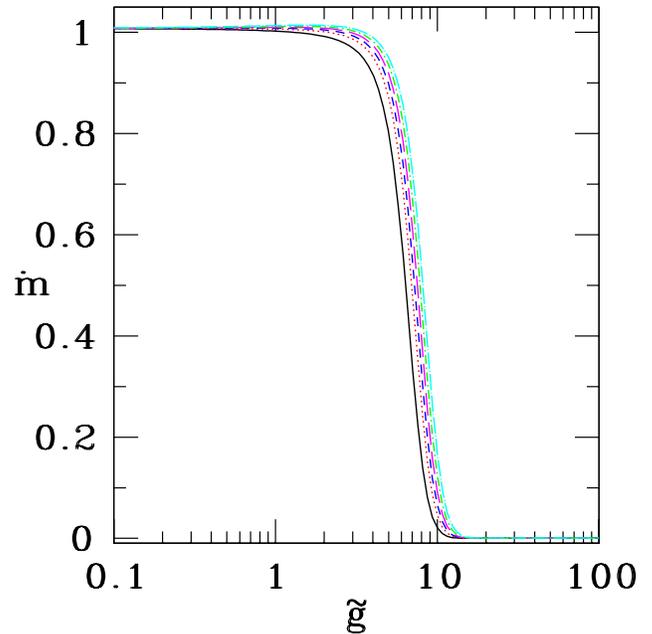}
\caption{Accretion rate $\dot m$ as a function of the tidal torque
parameter $\tilde{g}$ for various values of the viscosity power-law index
$ -1 \leq n \leq 2$. Results are shown for binary orbital radius $a = 100 M$,
disk thickness $h/r = 0.1$ in the vicinity of
the disk edge at $r \approx a$ (see Eq.~\ref{eq:torqtid}),
disk outer radius $r_{\rm out} = 10^5 M$ and
disk ISCO at $r_{\rm isco} = 6 M$. Solid (black) line represents
the case for $n=-1$, dotted (red) line for $n=0$, dashed (blue) line
for $n=1/2$, long-dashed (magenta) line for $n=1$, dot-dashed (green) line
for $n=3/2$ and dot-long-dashed (cyan) line for $n=2$. 
}
\label{fig:mdotvsg}
\end{figure}

\begin{figure}
\begin{center}
\includegraphics[width=9cm]{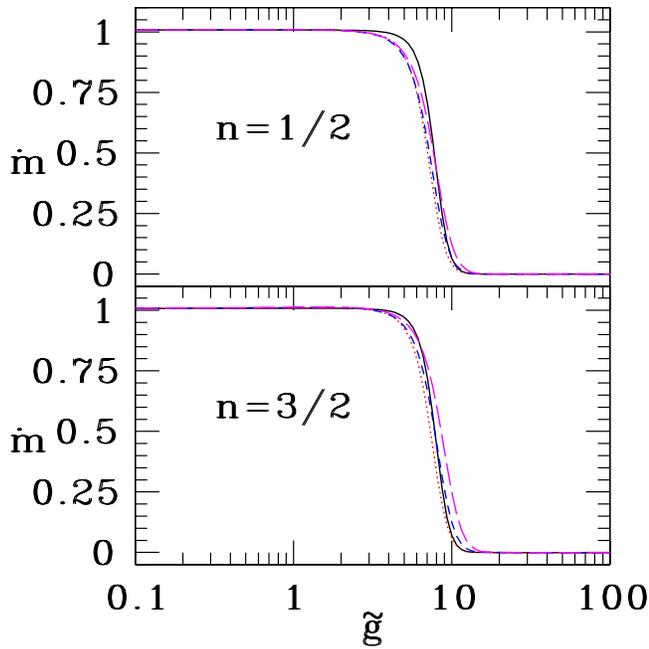}
\end{center}
\caption{Accretion rate $\dot{m}$ as a function of the tidal torque 
parameter $\tilde{g}$ for various values of disk thickness $h/r$. 
Results are shown for the viscosity power-law index $n=1/2$ ({\it top}) and
$n=3/2$ ({\it bottom}),  with binary orbital radius
$a = 100 M$, disk outer radius $r_{\rm out} = 10^5 M$ and
disk ISCO at $r_{\rm isco} = 6 M$. Solid (black) line represents
the case for $h/r=0.01$, dotted (red) line for $h/r=0.05$, dashed (blue) line
for $h/r=0.1$, and long-dashed (magenta) line for $h/r=0.15$. 
}
\label{fig:mdotvshovr}
\end{figure}

\begin{figure}
\includegraphics[width=8cm,height=6cm]{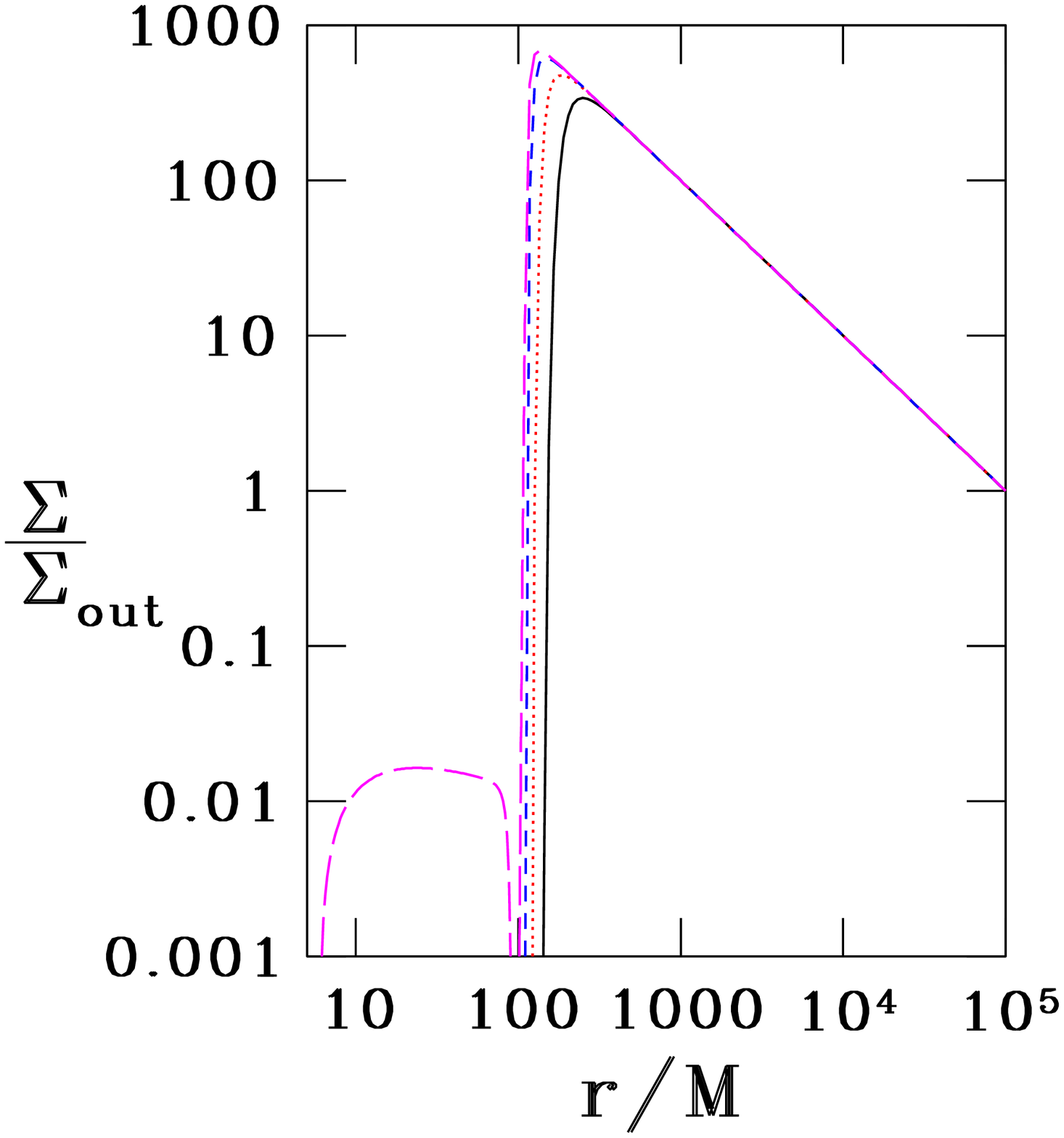}
\includegraphics[width=8cm,height=6cm]{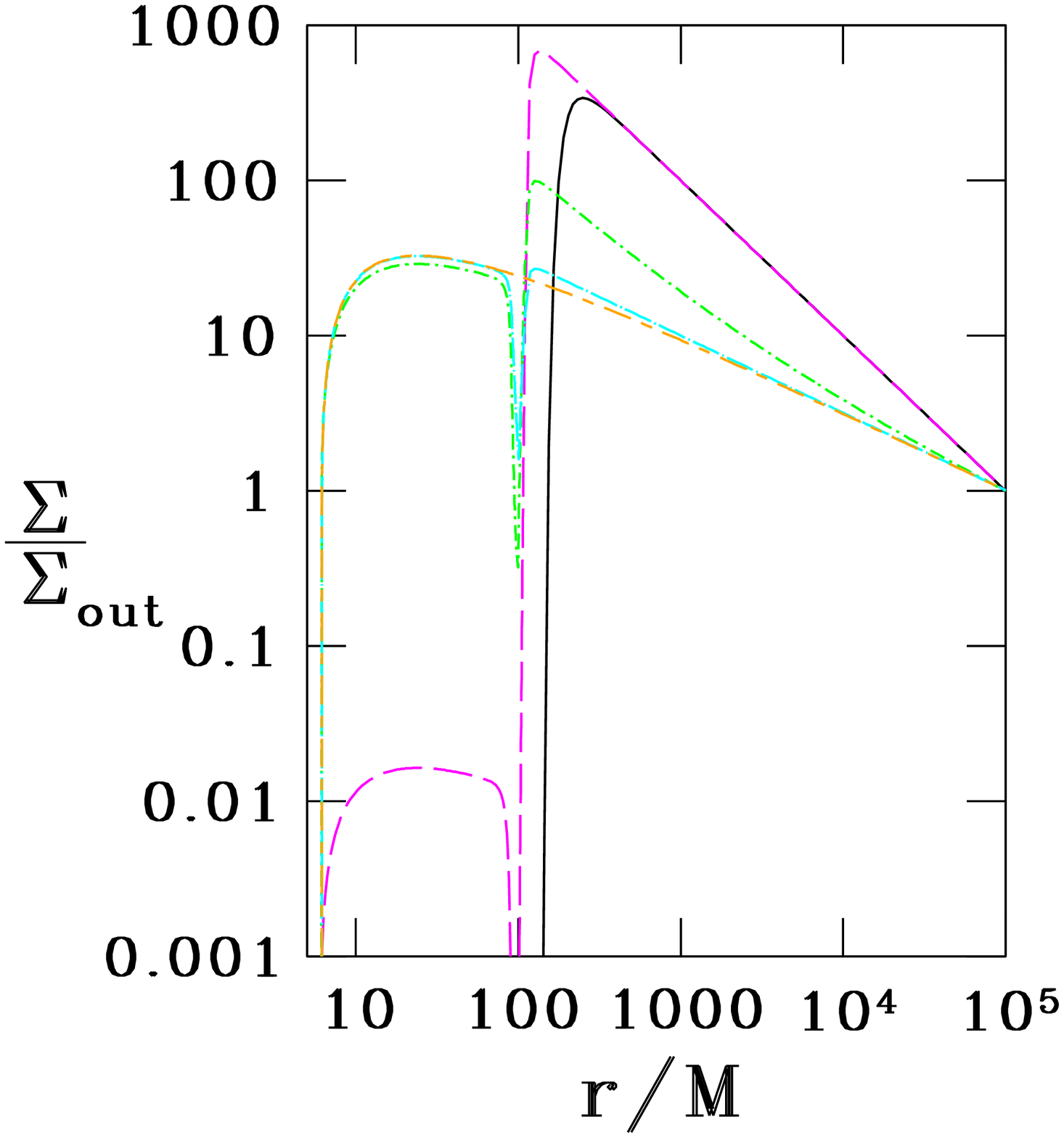}
\includegraphics[width=8cm,height=6cm]{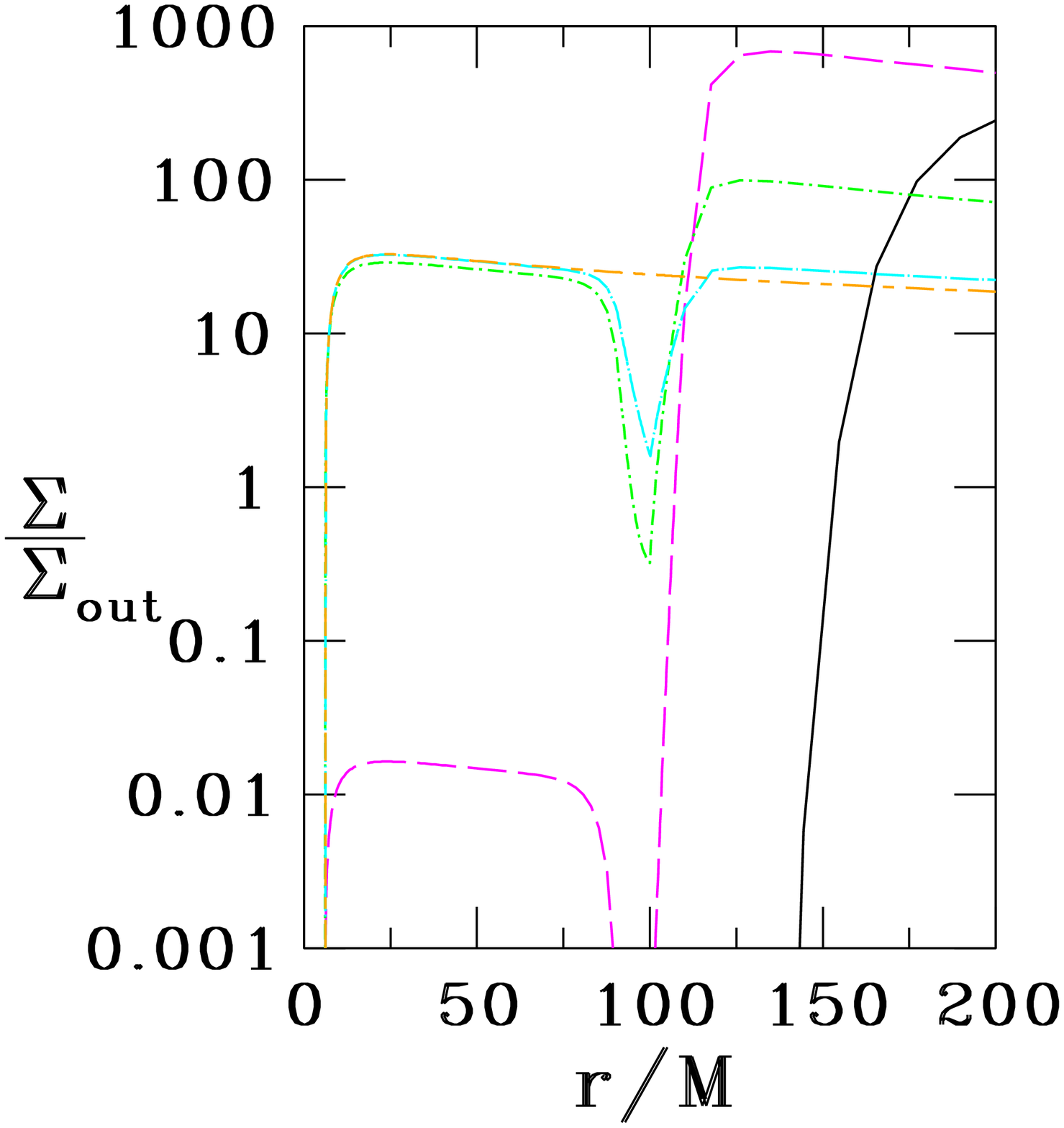}
\caption{The surface density profile $\Sigma$ plotted for different
values of the tidal torque parameter $\tilde{g}$. All results assume
$h/r = 0.1$, $n = 1/2$, $a = 100 M$, $r_{\rm out} = 10^5 M$ and
$r_{\rm isco} = 6 M$.  {\it Top}: $\tilde{g}=5000$ (black solid line),
$\tilde{g}=500$ (red dotted line),
$\tilde{g}=50$ (blue short-dash line) and $\tilde{g}=15$ (magenta long-dash line).
{\it Middle}: $\tilde{g}=5000$ (black solid), $\tilde{g}=15$
(magenta long-dash line), $\tilde{g}=5$
(green dot-short-dash line), $\tilde{g}=2.5$ (cyan dot-long-dash line), and
$\tilde{g}=0$ (orange long-dash-short-dash line).
{\it Bottom}: Same as the middle figure but plotted in linear scale in $r/M$.
}
\label{fig:sigvsr}
\end{figure}

Here we adopt the numerical recipe outlined in Section ~\ref{subsec:gensoln} to 
evaluate the quasistationary density profile for different 
disk parameters. We adopt a power law viscosity $\nu(r) \propto r^n$. 
We note that in their
vertically-integrated hydrodynamical simulations ref.~\cite{MacM08}
adopts $n=1/2$ for the value of the viscosity
power-law index, while in their integrations of the radial secular evolution
equations ref.~\cite{ArmN02} adopts $n=3/2$, while ref.~\cite{TanM10}
considers $n = 0.4$ and $n=0.065$.
In principle, the viscosity dependence, as well as the disk 
thickness, can be determined self-consistently once a viscosity 
law is adopted (e.g.\ an $\alpha$-disk prescription). But due to the 
uncertainty in this law, we treat these as free parameters 
and postpone a self-consistent calculation 
for a future investigation.

In Fig.~{\ref{fig:mdotvsg} we show the dependence of
the accretion rate $\dot m = \dot M/(3 \pi \nu_{\rm out} \Sigma_{\rm out})$ on the strength of the tidal torque parameter $\tilde{g}$
defined in Eq.~(\ref{eq:gtilde}) for fixed binary 
radius $a=100M$, disk thickness $h/r=0.1$ and disk outer-radius 
$r_{\rm out}=10^5M$. In Fig.~\ref{fig:mdotvshovr}, we show the $\tilde{g}$
dependence of $\dot{m}$ for $n=1/2$ and $n=3/2$ and for various values of 
disk thickness $h/r$ between 0.01 and 0.15, keeping 
other parameters fixed. 

The key result emerging from these two figures is that for small
tidal torques $\tilde{g} \lesssim \tilde{g}_c$ the accretion rate is very 
close to the rate onto an isolated black hole for the same asymptotic 
disk parameters,
while for strong torques $\tilde{g} \gtrsim \tilde{g}_c$, it is sharply reduced. 
This behavior is predicted from the simple analysis in Secs.~\ref{sec:Notidal} 
and \ref{sec:strongtid}. 
Since the parameter $\tilde{g}$ varies as $q^2$, the accretion rate 
thus decreases as the binary mass ratio increases. If we define $\tilde{g}_c$ 
to be the value of $\tilde{g}$ so that $\dot{m}=0.5$, we find $\tilde{g}_c$ lies 
between 5 and 10 for all the cases considered here, which is consistent with the 
value 6 estimated in Sec.~\ref{sec:strongtid}. 



In Fig.~\ref{fig:sigvsr} we show the surface density profile
for different tidal torque strengths $\tilde{g}$, fixing the viscosity index to 
$n=1/2$, disk thickness $h/r=0.1$, binary separation $a=100M$ and 
$r_{\rm out}=10^5M$. It is evident from the figure that when the tidal torque
is large and $\tilde{g} \gtrsim 5$, 
an appreciable gap develops in the 
density profile at the orbital radius of the secondary near $r=a$.
Moreover, the gap widens and the density inside the gap falls off sharply 
as $\tilde{g} \gtrsim 15$. By contrast, 
when the tidal torque is small and 
$\tilde{g} \lesssim 5$, the gap shrinks and the density profile 
approaches the stationary profile around a single black hole of the same mass
as the binary (Eq.~\ref{eq:signotorque}). We thus find explicitly
that the stationary solution yields an {\it inner} as well as an {\it outer} 
disk for sufficiently weak tidal torques (e.g., binaries with small $q$) and 
hardly any inner disk for strong tidal torques (e.g., binaries with
large $q$), as predicted in Sec.~\ref{sec:strongtid}. Correlating the values of
$\dot{m}$ with the density profiles for $\Sigma$ for each value of $\tilde{g}$
shows that the asymptotic profiles plotted in Fig.~\ref{fig:sigvsr} agree with
those derived in Eq.~(\ref{eq:sigasym}). 

We note that for large $\tilde{g}$ when the tidal torque is very 
strong, there is essentially no inner disk and the gas is unable cross 
the binary orbit. The surface density 
$\Sigma$ plummets near $r=a$. Hence the disk profile can also be obtained 
by imposing the boundary condition $\Sigma=0$ at $r=a$ for sufficiently large 
$\tilde{g}$. 
We have verified that for large $\tilde{g}$, imposing the boundary 
condition at $r=a$ gives the same disk profile as that obtained by 
imposing the boundary condition at $r=r_{\rm ISCO}$, confirming 
our intuition. Since we do not know a priori exactly how high 
$\tilde{g}$ must be for this approximation to apply, we must 
impose the true physical boundary condition at $r=r_{\rm ISCO}$ 
in general.

\section{Electromagnetic Spectrum}

\begin{table*}
\caption{Properties of disks around a binary black hole at the decoupling 
separation. The mass of the primary black hole is set to $M=10^8 M_\odot$, 
outer disk radius $r_{\rm out}=10^3 M$, surface density 
$\Sigma_{\rm out}=1.5\times 10^4 {\rm g\, cm}^{-2}$ and the Eddington 
parameter $\gamma=0.1$.}
\begin{tabular}{cccccccccc}
\hline
 Case & $q$ & $\tilde{g}$ & $t_{\rm orb}$(days) & $t_{GW}$(yrs) & $a_d/M$ & $M_{\rm disk} (M_\odot)$ 
& $\dot{m}$ & $\dot{M} (M_\odot {\rm \, yr}^{-1})$ & $L ({\rm erg\, s}^{-1})$ \\
\hline
\hline
 a & 0 & 0 & -- & -- & -- & 660 & 1.08 & 0.29 & $1.3\times 10^{45}$ \\
 b & $2\times 10^{-3}$ & 1.6 & 1.6 & 16 & 12.6 & 660 & 1.07 & 0.29 & $1.5\times 10^{45}$ \\
 c & $3\times 10^{-3}$ & 3.5 & 2.0 & 20 & 14.8 & 700 & 0.953 & 0.26 & $3.0\times 10^{45}$ \\
 d & $4\times 10^{-3}$ & 6.3 & 2.4 & 23 & 16.6 & 890 & 0.381 & 0.10 & $9.1\times 10^{45}$ \\
 e & $5\times 10^{-3}$ & 9.8 & 2.8 & 27 & 18.2 & 1000 & 0.0197 & $5.3\times 10^{-3}$ & $1.1\times 10^{46}$ \\
 f & $10^{-2}$ & 39 & 4.2 & 40 & 23.9 & 1000 & 0 & 0 & $6.5\times 10^{45}$ \\
 g & $10^{-1}$ & 3900 & 15 & 145 & 56.0 & 910 & 0 & 0 & $1.4\times 10^{45}$ \\
\hline
\end{tabular}
\label{tab:cases}
\end{table*}

Here we compute the (approximate) electromagnetic spectra from
stationary disks for some representative, astrophysically plausible systems.
We are particularly
interested in probing disks at the time of decoupling and
immersed in gaseous regions that will
yield stationary post-merger accretion rates onto the {\it remnant} black 
hole near the Eddington limit. After decoupling, our stationary disk solution 
is no longer valid. 
To specify a model we first choose the parameters $M, q, a/M,
r_{\rm out}/M \gg 1$ and $n$. 
We determine the decoupling separation $a_d$ by the condition 
$t_{\rm GW}(a_d) = \beta t_{\rm vis}(2a_d)$ (setting $\beta=0.1$), 
which yields 
\begin{equation}
\left(\frac{a_d}{M}\right)^2 = \frac{128 M \beta \zeta}{15 \nu(2a_d)}
= \frac{128 M \beta \zeta}{15 \nu_{\rm out}} \left( \frac{r_{\rm out}}{2a_d}\right)^n
\end{equation}
or 
\begin{equation}
  \frac{a_d}{M} = \left[ \frac{128M\beta \zeta}{15 \nu_{\rm out}} 
\left(\frac{r_{\rm out}}{2M}\right)^n\right]^{1/(n+2)}.
\label{eq:ad2}
\end{equation}
We next set $\bar{h}=h/r =0.1$ for the disk
thickness near $r=a$. To establish the scale for the density and disk size in physical units we fix 
$\Sigma_{\rm out}$ and $r_{\rm out}$, which determines the disk mass $M_{\rm disk}$.  
Finally, we determine $\nu_{\rm out}$ by specifying the accretion 
rate onto the black hole remnant, 
$\dot{M}_{\rm rem} = 3 \pi \nu_{\rm out} \Sigma_{\rm out}$ 
(see Eq.~\ref{eq:mdotnotorque2} in the limit $r_{\rm isco}/r_{\rm out} \ll 1$) 
to be a fraction $\gamma$ of the Eddington value, whereby
\begin{equation}
\gamma \equiv \frac{\dot{M}_{\rm rem}}{\dot{M}_{\rm Edd}} =
\frac{3 \pi \nu_{\rm out} \Sigma_{\rm out}}{\dot{M}_{\rm Edd}},
\end{equation}
which gives
\begin{equation}
\nu_{\rm out} = \frac{16 \gamma M m_p}{\sigma_T \Sigma_{\rm out}}.
\end{equation}
The above choices determine the torque parameters $g$ and $\tilde{g}$ 
according to Eqs.~(\ref{eq:g}) and (\ref{eq:gtilde}), setting $f=0.01$.
These choices specify a unique, albeit approximate, model for a 
stationary Newtonian disk about a binary black 
hole system and determine the disk surface density and
the gas accretion rate.
Given such a model, the electromagnetic spectrum can be computed following 
the prescription outlined in Section~\ref{sec:em1} above.

\subsection{Numerical Integrations}

\begin{figure}
\includegraphics[width=9cm]{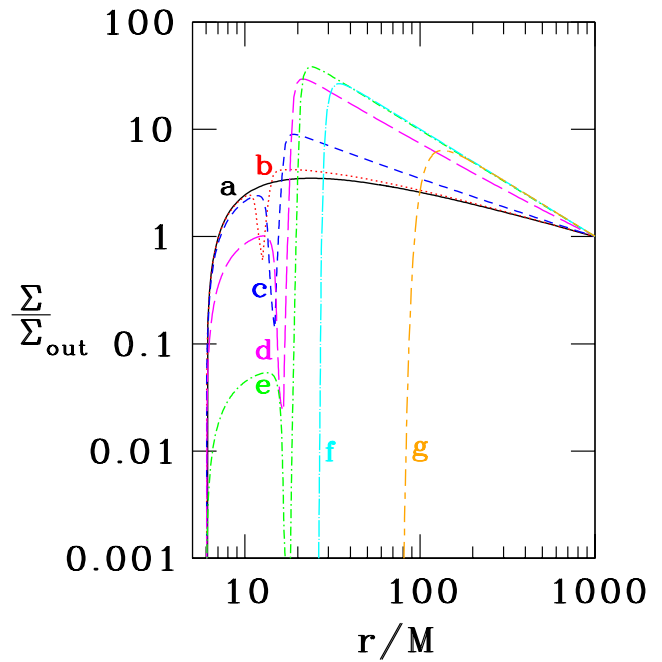}
\caption{Surface density of disks around a binary black hole just before decoupling.
Shown here are cases for the binary mass ratios
(a) $q=0$ (black solid line), (b) $q=2\times 10^{-3}$ (red dotted line),
(c) $q=3\times 10^{-3}$ (blue short-dash line), (d) $q=4\times 10^{-3}$
(magenta long-dashed line), (e) $q=5\times 10^{-3}$ (green dot-short-dash
line), (f) $q=10^{-2}$ (cyan dot-long-dash line), and (g) $q=10^{-1}$
(orange long-dash-short-dash line).
}
\label{fig:sigma_d}
\end{figure}

\begin{figure}
\includegraphics[width=9cm]{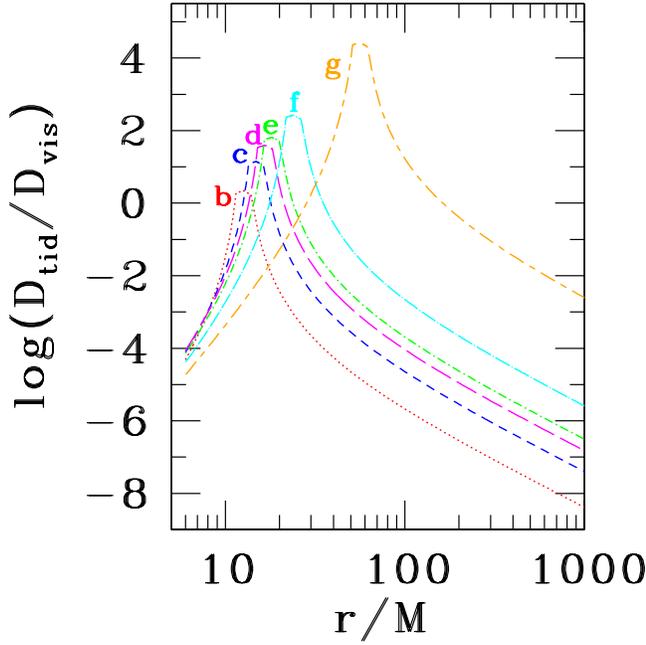}
\caption{Ratio of tidal heating to viscous heating $D_{\rm tid}/D_{\rm vis}$
as a function of radius for disks around a binary black hole just before decoupling
for binary ratios (b) $q=2\times 10^{-3}$
(red dotted line),
(c) $q=3\times 10^{-3}$ (blue short-dash line), (d) $q=4\times 10^{-3}$
(magenta long-dashed line), (e) $q=5\times 10^{-3}$ (green dot-short-dash
line), (f) $q=10^{-2}$ (cyan dot-long-dash line), and (g) $q=10^{-1}$
(orange long-dash-short-dash line).
}
\label{fig:Dtid_Dvis}
\end{figure}

\begin{figure}
\includegraphics[width=9cm]{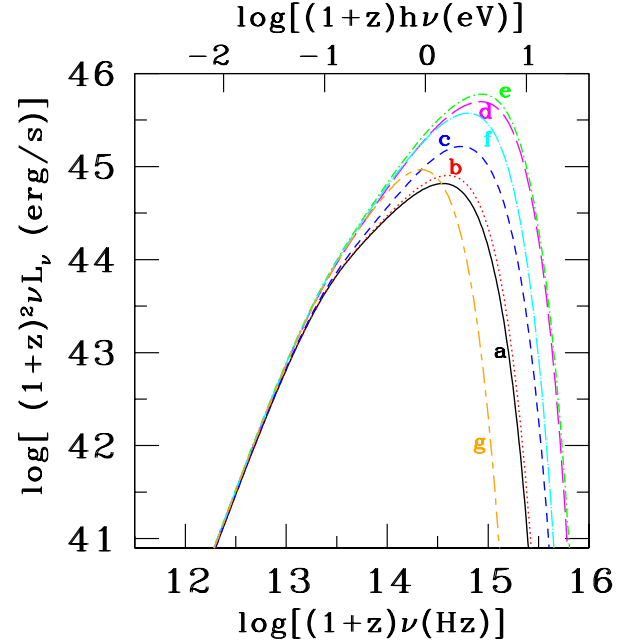}
\caption{Electromagnetic spectrum of disks around a binary black hole at redshift $z$
just before decoupling. Shown here are cases for the binary mass ratios
(a) $q=0$ (black solid line), (b) $q=2\times 10^{-3}$ (red dotted line),
(c) $q=3\times 10^{-3}$ (blue short-dash line), (d) $q=4\times 10^{-3}$
(magenta long-dashed line), (e) $q=5\times 10^{-3}$ (green dot-short-dash
line), (f) $q=10^{-2}$ (cyan dot-long-dash line), and (g) $q=10^{-1}$
(orange long-dash-short-dash line).}
\label{fig:nuLnu}
\end{figure}

We illustrate the calculation by considering a binary black hole with
a $10^8 M_{\odot}$ primary.  We take the viscosity power-law index to be
$n=0.5$, and set $r_{\rm out} = 10^3M$, $\gamma=0.1$, 
and $\Sigma_{\rm out}=1.5\times 10^4 {\rm g\, cm}^{-2}$. The decoupling 
separation $a_d$ is calculated from Eq.~(\ref{eq:ad2}). Our choice of 
parameters gives
$M_{\rm disk} \sim 10^3 M_\odot$ for $0 \leq q \leq 0.1$
(see Table~\ref{tab:cases}), which is much smaller than the mass of the 
secondary BH for $q>10^{-3}$. These values for the disk density and 
mass are comparable to those considered in~\cite{ArmN02,ChaSMQ10}.
We evaluate the disk profiles and electromagnetic 
structure for different choices of mass ratio $q \leq 0.1$. 
Table~\ref{tab:cases} summarizes our results. Scaling behavior for different 
choices of binary and disk parameters is presented in Appendix~\ref{app:scaling}.
We see that the 
decoupling separation $a_d\propto \zeta^{0.4}$ 
[see Eq.~(\ref{eq:ad2})] increases from $12.6M$ for $q=2\times 10^{-3}$ to 
$56M$ for $q=0.1$. The accretion rate drops rapidly towards 0 when 
$q>4 \times 10^{-3}$, which corresponds to $\tilde{g} > 6.3$, and is 
consistent with Fig.~\ref{fig:mdotvsg}. 

Figure~\ref{fig:sigma_d} shows the surface density of the disk 
for different values of $q$. As in Fig.~\ref{fig:sigvsr}, for small $q$, there 
is a dip near $r=a$ in the density profile, but an inner disk of substantial 
density is present. When $q>4\times 10^{-3}$, 
the accretion rate drops to a small value and $\Sigma$ falls sharply 
near $r=a$, leaving essentially no inner disk.
We also see that the density in the asymptotic region changes from 
$r^{-1}$ for small $\dot{m}$ (large $q$) to $r^{-1/2}$ for substantial 
$\dot{m}$ (small $q$), as 
predicted by Eq.~(\ref{eq:sigasym}). For a fixed $\Sigma_{\rm out}$, 
the steeper decrease of density in the small accretion cases 
($q>4\times 10^{-3}$) give larger $\Sigma$ in the asymptotic region, 
which results in a larger disk mass as shown in Table~\ref{tab:cases}. 
However, as $q$ increases further, the decoupling radius $a_d$ moves 
to a larger radius so that the inner radius of the disk moves out as well. 
This behavior leads to the small decrease in disk mass for $q>10^{-2}$ as 
shown in Table~\ref{tab:cases}.

Figure~\ref{fig:Dtid_Dvis} shows the ratio of tidal heating to viscous heating
$D_{\rm tid}/D_{\rm vis}$ as a function of radius for each case. In all 
cases with $q \neq 0$,
the effect of tidal heating dominates over the viscous heating at radius
$r \sim a$, but decreases rapidly both at larger and smaller radii. This result  
is expected since the tidal torque is created by the gravitational force 
of the secondary black hole, which decreases with the distance from the 
secondary. 
It can be proven easily from Eqs.~(\ref{diss}), (\ref{eq:Dtid}), 
(\ref{eq:AN}) and
(\ref{eq:torqtid}) that $D_{\rm tid}/D_{\rm vis} \sim r^3 |\Lambda|/\nu$.
Hence $D_{\rm tid}/D_{\rm vis}$ has a peak near $r=a$,
$D_{\rm tid}/D_{\rm vis}\sim r^{3/2}/(r-a)^4$ for $r\gtrsim a$ and
$D_{\rm tid}/D_{\rm vis}\sim r^{11/2}/(r-a)^4$ for $r \lesssim a$.

Figure~\ref{fig:nuLnu} shows the electromagnetic spectrum of the disk for each value of $q$. 
We see that the luminosity increases substantially at first as $q$ increases 
above zero and the accretion rate $\dot{m}$ 
drops. This is due to the large increase in $\Sigma$ as shown in Fig.~\ref{fig:sigma_d}. The 
peak frequency also shifts to a higher value because the enhanced viscous and tidal dissipation 
give rise to higher temperatures in the disk. For $q \gtrsim 5\times 10^{-3}$, however, 
the tidal torque of the binary becomes large enough to expel the disk material inside its 
orbit and shut off the accretion. As a result, the edge of the disk moves to a larger radius where 
the temperature is lower. This effect shifts the peak frequency back to a lower value. 
Since a large 
fraction of the electromagnetic radiation comes from the innermost part 
of the disk where the temperature 
is high, the disk luminosity drops when $q \gtrsim 5\times 10^{-3}$ 
(see Table~\ref{tab:cases}), as the inner edge of the disk moves to a larger radius. We see from 
Table~\ref{tab:cases} that the total luminosity of the disk for $q=0.1$ 
drops to about the same value as the standard Shakura-Sunyaev disk. 

Most noteworthy is the fact that even as the gas accretion rate is shut off by the companion, 
tidal heating serves to maintain a high luminosity from the disk, comparable in magnitude 
to the luminosity from a single black hole. For high $q$ cases in which there is no inner 
disk, the luminosity plummets after decoupling over a timescale $t_{\rm GW}$, as the inspiral 
accelerates and tidal heating drops. Later on the luminosity from the disk about 
the remnant then increases on a slow viscous timescale, as the outer disk diffuses 
inward~\cite{MilP05,Sha10,TanM10}. This luminosity achieves values comparable to those 
prior to decoupling, but at higher frequency, since the main radiation region outside the 
inner edge of the disk moves inward from $a_d$ to $r_{\rm isco}$, where temperatures become 
hotter. For low $q$ cases in which there is a substantial inner disk the change in luminosity 
and the frequency spectrum following decoupling is much less pronounced.

Periodicities in the luminosity are expected on an orbital timescale 
during the inspiral. This timescale is on the order of days at 
decoupling for the cases considered here (see Table~\ref{tab:cases}).

Figure~\ref{fig:nuLnu} indicates that most of the radiation emitted by the disk 
at decoupling is in visible and near infrared wavelengths. For a source 
at redshift $z=1$, the apparent magnitude is about 19 and radiated in
infrared wavelength. Assuming the disk 
is not obscured by interstellar dust, it could be observable by  
the James Webb Space Telescope (JWST) and Large Synoptic Survey Telescope (LSST).

\section{Summary}

We have studied the effects of tidal torque on a 
quasistationary disk around a binary black hole with small binary 
mass ratio before decoupling. We find that the density profile and 
accretion rate is sensitive to the mass ratio $q$ through the 
tidal torque parameter $\tilde{g}$ defined by Eq.~(\ref{eq:gtilde}).
For small $\tilde{g}$, 
the density profile and accretion rate are slightly modified from 
the standard Shakura-Sunyaev disk. Specifically, the density inside 
the binary orbit ($r<a$) is smaller than the standard Shakura-Sunyaev 
disk but is about the same in the outer region. A substantial drop in 
density is seen near $r=a$ due to the strong tidal torque in that region. 
As $\tilde{g}$ increases, the tidal torque becomes stronger and the density in 
the inner disk, as well as the accretion rate, decrease further. On the 
other hand, the density in the outer disk increases and a gap (inside of which 
the density is much smaller) develops between the inner and outer disk. 
The gap widens with increasing $\tilde{g}$. When $\tilde{g}$ exceeds a critical 
value $\tilde{g}_c$, 
the tidal torque is strong enough to effectively expel the material 
inside and near the binary orbit and practically shuts off accretion. 
The asymptotic density profile also changes [see Eq.~(\ref{eq:sigasym})]. 
For a power law viscosity $\nu \propto r^n$, 
the critical value is $\tilde{g}_c \sim 7$ and depends weakly 
on the index $n$ and disk thickness $h/r$ 
(see Figs.~\ref{fig:mdotvsg} and \ref{fig:mdotvshovr}). 

We compute the luminosity and electromagnetic spectrum 
emitted by the disk due to viscous heating and tidal dissipation at the 
time of decoupling. The results are summarized in Table~\ref{tab:cases} and 
Fig.~\ref{fig:nuLnu} for disks surrounding a primary black hole of mass 
$M=10^8M_\odot$, outer disk radius $r=10^3M$, surface density 
$\Sigma_{\rm out}=1.5\times 10^4 {\rm g\, cm}^{-2}$, and the Eddington 
parameter $\gamma=0.1$. The disk luminosity at first increases 
with $q$. This is caused by the increase in density of the outer disk. 
The radiation also shifts to higher frequencies, as the enhanced viscous and 
tidal heating give rise to higher disk temperatures. The disk luminosity 
reaches a maximum near $q=5\times 10^{-3}$, when the accretion is shut off by the strong 
tidal torques. When $q>5\times 10^{-3}$, the decoupling binary separation $a_d$ moves to 
a larger radius. The strong tidal torque expels disk material inside and near the binary orbit, 
causing the radiation region to move to larger radii, where the temperatures 
are lower. The disk luminosity drops and the radiation shifts to 
a lower frequency. We note that even as the gas accretion rate is shut off by the companion, 
tidal heating serves to maintain a high luminosity from the disk, comparable in magnitude 
to the luminosity from a single black hole. Most of the radiation is emitted in the near 
infrared and visible wavelength and may exhibit periodicities on the 
binary orbital timescale. For a source at redshift $z=1$, the apparent magnitude is 
about 19 (provided the disk is not obscured by the interstellar dust) and could be detected by 
JWST and LSST.

{\it Acknowledgments}: It is a pleasure to thank 
C. Gammie for useful discussions. This paper was supported in part by NSF
Grants PHY06-50377 and PHY09-63136 and NASA Grants NNX07AG96G and NNX10A173G 
to the University of Illinois at Urbana-Champaign.

\appendix 

\section{Scaling in disk radiation}
\label{app:scaling}
In this appendix, we rewrite the equations in Sec.~\ref{sec:em1} in terms of the 
nondimensional variables introduced in Sec.~\ref{sec:nondim} and then establish 
scaling relations in terms of the mass of the primary black hole $M$ and the 
Eddington parameter $\gamma$.

We first rewrite Eq.~(\ref{diss}) as 
\beq
D_{\rm vis}(t,r) 
  = \frac{9}{8}\frac{\nu_{\rm out} \Sigma_{\rm out} M}{r_{\rm out}^3} \left( \frac{\ybar}{s^7}\right) \ .
\label{eq:Dvis}
\eeq
The torque function $\Lambda$ can be written as 
\beq
\Lambda = \frac{3\nu_{\rm out} \sqrt{M}}{4r_{\rm out}^{3/2}s^2} f(s) \nubar(s). 
\eeq
Hence the rate of change in binary binding energy is [see Eq.~(\ref{eq:Edottid})]
\beqn
  \dot{E}_{\rm tid} &=& \frac{3\pi M \sqrt{r_{\rm out}} \nu_{\rm out} \Sigma_{\rm out}}{a^{3/2}} 
\int_{s_2}^1 f(s) \ybar(s) ds \cr
&=& \frac{3\pi M \sqrt{r_{\rm out}} \nu_{\rm out} \Sigma_{\rm out}}{a^{3/2}} [ 1-\dot{m}(1-s_2)] \ .
\eeqn
The tidal torque density Eq.~(\ref{eq:AN}) becomes 
\beq
\frac{\partial T_{\rm tid}}{\partial r}  
 = \frac{3\pi \Sigma_{\rm out} \nu_{\rm out}}{2}\sqrt{\frac{M}{r_{\rm out}}} \, \frac{f(s)\ybar(s)}{s} ,
\eeq
Consider the expression
\beqn
  \int_{r_{\rm isco}}^{r_{\rm out}} dr |dT_{\rm tid}/dr| &=& 2r_{\rm out} \int_{s_2}^1 s |dT_{\rm tid}/dr| ds \cr
 &=& 3\pi \Sigma_{\rm out} \nu_{\rm out} \sqrt{Mr_{\rm out}} \left[ -\int_{s_2}^{s_1} f(s) \ybar(s) ds 
\right. \cr 
&& \left. + \int_{s_1}^1 f(s) \ybar(s) ds \right] \cr 
 &=& 3\pi \Sigma_{\rm out} \nu_{\rm out} \sqrt{Mr_{\rm out}}\times \cr
&& [ 1-2\ybar(s_1)+\dot{m}(2s_1-s_2-1)] .
\eeqn
Hence Eq.~(\ref{eq:Dtid}) becomes 
\beq
  D_{\rm tid} = \frac{3M\nu_{\rm out} \Sigma_{\rm out} [1-\dot{m}(1-s_2)]}{8(r_{\rm out}a)^{3/2} 
[1-2\ybar(s_1)+\dot{m}(2s_1-s_2-1)]} \frac{ |f(s)|\ybar(s)}{s^3} .
\eeq
The total dissipation rate is 
\beqn
  D &=& D_{\rm tid} + D_{\rm vis} \cr \cr
  &=& \frac{3M\nu_{\rm out}\Sigma_{\rm out}}{8r_{\rm out}^3} \bar{D}(s) ,
\eeqn 
where 
\beq
\bar{D}(s) = \left[ \frac{3\ybar}{s^7} + 
 \frac{1-\dot{m}(1-s_2)}{1-2\ybar(s_1)+\dot{m}(2s_1-s_2-1)} 
\frac{|f(s)|\ybar(s)}{s^3 s_1^3}\right] . \ \ 
\eeq
The temperature ratio is given by 
\beq
  \left(\frac{T_*}{T_s}\right)^4 = \frac{D}{3M\dot{M}_{\rm rem}/8\pi r_{\rm isco}^3} 
= \frac{3}{s_2^6} \bar{D}(s) .
\eeq

Gathering all the relevant formulae, the accretion rate, luminosity, and disk mass 
can be expressed in cgs units as follows: 
\beq
 \dot{M} = 0.27 M_\odot {\rm \, yr}^{-1} \left( \frac{\gamma}{0.1}\right) 
\left( \frac{M}{10^8 M_\odot} \right) \dot{m} ,
\eeq
\beq
 L_\nu 
 = 1.99\times 10^{30} {\rm erg} \left( \frac{\gamma}{0.1}\right)^{3/4} 
\left( \frac{M}{10^8 M_\odot}\right)^{5/4} f^*(\nu/\nu_*) ,
\eeq
\beq
  \nu_* = 2.91\times 10^{14} \left(\frac{\gamma}{0.1}\right)^{1/4}  
\left(\frac{M}{10^8 M_\odot}\right)^{-1/4} {\rm Hz} ,
\eeq
\beqn
  L &=& 1.26\times 10^{45} \bar{L} 
\left(\frac{\gamma}{0.1}\right) \left( \frac{M}{10^8 M_\odot} 
\right) {\rm erg \ s}^{-1} , \\ 
 \bar{L} &\equiv& 2s_2^2 \int_{s_2}^1 \bar{D}(s) ds , 
\eeqn
and 
\beqn
  M_{\rm disk} 
 &=& 2.06\times 10^3 M_\odot \left(\frac{\Sigma}{1.5\times 10^4{\rm g\, cm}^{-2}}\right)
\left( \frac{r_{\rm out}}{10^3M}\right)^2 \times \cr 
&&\left( \frac{M}{10^8 M_\odot}\right)^2 
 \int_{s_2}^1 s^{2-2n} \ybar ds .
\eeqn

\bibliography{paper}

\begin{thebibliography}{10}
\bibitem{DYN}%
  \BibitemOpen
  \bibinfo {note} {{There are exceptions, as in the case of nearly equal-mass
  binaries at merger, when abrupt mass loss and recoil due to gravitational
  wave emission occurs ~\cite{BodP07,CorHM09,RosLAPK09,OneMBRS09} }}%
  \bibAnnoteFile{NoStop}{DYN}%
\bibitem{FREQ}%
  \BibitemOpen
  \bibinfo {note} {{From the context it should not be difficult to distinguish
  the photon frequency $\nu$ from the viscosity $\nu$.}}%
  \bibAnnoteFile{Stop}{FREQ}%
\end{thebibliography}
\end{document}